\documentclass[aps,prl,preprint,showpacs,superscriptaddress,amssymb,nobibnotes]{revtex4-1}
\usepackage{graphicx}  
\usepackage{dcolumn}   
\usepackage{bm}        
\usepackage{amssymb}   
\usepackage{multirow}

\begin{document}


\title{Manipulating Magnetism at Organic/Ferromagnetic Interfaces by Fullerene-Induced Surface Reconstruction}


\author{Rui Pang}
\affiliation{Department of Physics, South University of Science and Technology of China, Shenzhen 518055, China}
\author{Xingqiang Shi}\email[]{shixq@sustc.edu.cn}
\affiliation{Department of Physics, South University of Science and Technology of China, Shenzhen 518055, China}
\author{Michel A. Van Hove}
\affiliation{Institute of Computational and Theoretical Studies and Department of Physics, Hong Kong Baptist University, Hong Kong, China}



\date{\today}

\begin{abstract}
Fullerenes have several advantages as potential materials for organic spintronics.
Through a theoretical first-principles study, we report that fullerene C$_{60}$ adsorption can induce a magnetic reconstruction in a Ni(111) surface and expose the merits of the reconstructed C$_{60}$/Ni(111) \emph{spinterface} for molecular spintronics applications.
Surface reconstruction drastically modifies the magnetic properties at both sides of the C$_{60}$/Ni interface.
Three outstanding properties of the reconstructed structure are revealed, which originate from reconstruction enhanced spin-split $\mathrm{\pi}$$-$d coupling between C$_{60}$ and Ni(111):
1) the C$_{60}$ spin polarization and conductance around the Fermi level are enhanced simultaneously, which can be important for read-head sensor miniaturization;
2) localized spin-polarized states appear in C$_{60}$ with a spin-filter functionality, and
3) magnetocrystalline anisotropic energy and exchange coupling in the outermost Ni layer are reduced enormously.
Surface reconstruction can be realized simply by controlling the annealing temperature in experiments.
\end{abstract}


\maketitle

\section{}

Non-magnetic/ferromagnetic interfaces of magnetic hard disk drives are crucial for their performance.
To achieve read-head sensor miniaturization, it is essential to minimize the resistance of the system to maintain an ideal data-transfer rate and signal-to-noise ratio;
at the same time, a large magnetoresistance(MR) ratio is decisive for optimum functionality of read-heads\cite{point12,point13}.
It is therefore desired to find a method that can simultaneously increase the MR ratio and the conductance of the device\cite{point1}.
Recently, tuning the properties at organic/ferromagnetic interfaces by aromatic molecules has attracted broad attention\cite{review_organic,molecular_spintronics1,molecular_spintronics2}.
Due to the $\pi-d$ hybridization, chemical adsorption of aromatic molecules on magnetic surfaces produces new spin-split hybridized states at the interface (called a \emph{spinterface}) \cite{spinterface,spinterface1,spinterface2,spinterface3,spinterface4,review_organic_interface1,review_organic_interface2,method1,method2,surface_exchange2}. These states can be used to produce  thermal-robust molecular spintronic devices\cite{viewpoint_spinterface}.
A challenge of this method is that chemical adsorption usually broadens the molecular orbitals near the metal Fermi level, which acts against the desired appearance of
spin-polarized and energy-concentrated states\cite{viewpoint_spinterface,molecularlike_orbital}.
Another possible usage of the organic/ferromagnetic interface is that the effective magnetic coupling of surface atoms can be modified by the adsorbate\cite{surface_exchange1,surface_exchange2,surface_exchange3,surface_exchange4}. Thus it is possible to create hard/soft composite magnetic structures by self-assembly in molecular adsorption to achieve desired applications in permanent magnets, recording media and spintronics\cite{magnetic_spring}.

Fullerenes and their derivatives are building blocks of potential high performance organic devices\cite{fullerene_spintronics,c60_mr1,c60_mr2,c60_mr3,c60_mr4,nic60ni}.
Meanwhile, it has been proven that C$_{60}$ adsorption can induce \emph{non-magnetic} metal surface reconstruction, i.e. rearrangement with different bonding of surface atoms\cite{c60_reconstruction_al,c60_reconstruction_cu}.
These reconstructions have decisive influences on the their charge transport properties\cite{transport_reconstruction}.
Thus, extending to \emph{magnetic} metal surfaces, one can expect that adsorption-induced reconstruction could also have significant effects on the spin transport properties of fullerene/ferromagnetic interface.
It is crucial to identify the existence and magnetic effects of reconstruction at these organic/ferromagnetic interfaces.
For C$_{60}$/Ni(111), height profile measurements with scanning tunneling microscopy (STM) show that C$_{60}$ can adsorb at different heights above the Ni(111) surface\cite{c60_reconstruction_ni},
which is a hint for surface reconstruction.
However, the reconstructed atomic structure  and the effects of reconstruction on interface magnetic properties has not been studied to our knowledge.

In this letter we investigate the geometric and magnetic properties of the C$_{60}$/Ni(111) interface by first-principles methods.
We show that the reconstructed structure is energetically favored over the unreconstructed one.
We demonstrate that, in comparison with the unreconstructed structure, the reconstructed one has the following superior properties:
1) the density of states (DOS) and the spin polarization of C$_{60}$ are enhanced simultaneously around the Fermi level;
2) the molecular spin-polarized states are concentrated in energy around the Fermi level;
3) the magnetic coupling and magnetocrystalline anisotropic energies (MAE) of atoms in the outermost substrate layer are significantly reduced.
The above changes in properties  show that one can significantly affect magnetism at the organic/ferromagnetic interface through surface reconstruction. This prediction could have further applications in molecular and organic spintronics.

Calculations were performed using the plane-wave-basis-set Vienna ab-initio simulation package (VASP)\cite{vasp1}. The Ni(111) surface was modeled by a five-layer-slab with a 4$\times$4 surface unit cell per C$_{60}$, which cell size is determined from electron diffraction experiments\cite{leed1,leed2}.
Projector augmented wave potentials\cite{vasp2} were employed with a kinetic energy cutoff of 500 eV and with a $K$-point sampling of 4$\times$4. For the exchange-correlation functional, the Perdew-Burke-Ernzerh of generalized gradient approximation was utilized\cite{gga}. The calculations of magnetic couplings and MAE utilized the Quantum Espresso package \cite{quantum_espresso} with equivalent computational parameters as in the VASP calculations.
The adsorption structures with and without reconstruction were selected by (a) symmetry- and size-matching between C$_{60}$ and the Ni(111) surface, and (b) the informations of the same and similar systems\cite{c60_reconstruction_cu,c60_reconstruction_ni,reconstruction_candidate1,kink,c60_pt,c60_cu,c60_structure} (details see the Supplemental Material\cite{SI}).

\begin{figure}
\includegraphics[width=0.9\textwidth]{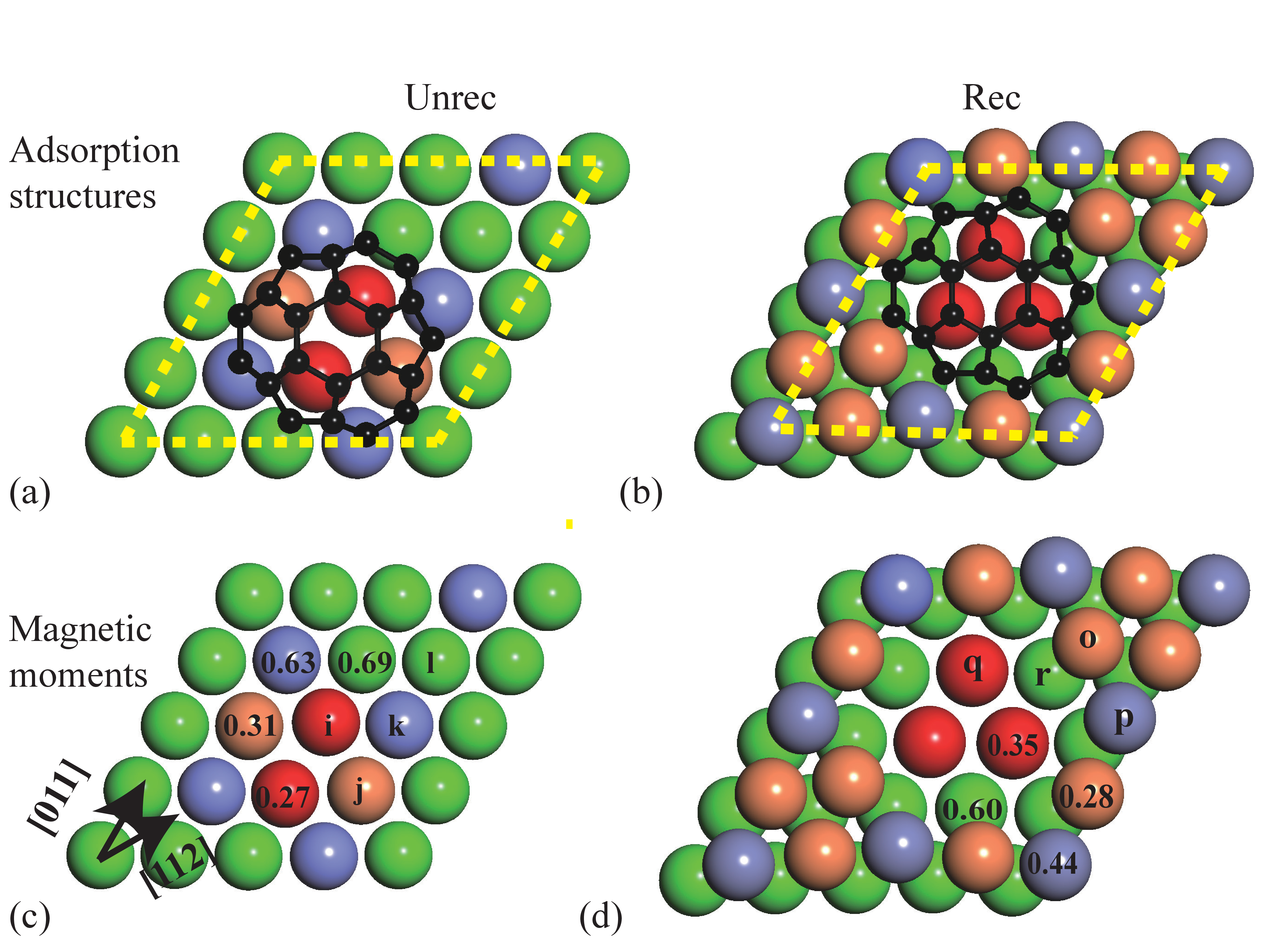}
\caption{\label{structure}
(a) and (b): Top views of the unreconstructed (Unrec) and reconstructed (Rec) structures of C$_{60}$/Ni(111); only the bottom part of C$_{60}$ and the top layer(s) of Ni are shown; the dashed lines outline a Ni(111)-(4$\times$4) surface unit cell. Different types of surface atoms are labeled with different colors classified by their magnetic moments and their distances from the C$_{60}$ carbon atoms.
(c) and (d): Magnetic moments (in $\mu_{\text{B}}$) of surface atoms in Unrec and Rec structures.}
\end{figure}

Fig.~\ref{structure} shows the most stable structures for both the reconstructed and unreconstructed cases (Rec and Unrec, resp.). Their stabilities, i.e., adsorption energies corrected for vacancy formation energies, are $-2.13$ eV for Unrec and $-3.41$ eV for Rec (see Supplemental Material\cite{SI}).

As shown in Fig.~\ref{structure}a and b, C$_{60}$ binds with Ni through a C$_{6}$ hexagon parallel to the Ni(111) surface.
In Unrec, the centre of gravity of C$_{60}$ locates at a bridge site of the surface. Six C$-$Ni bonds are formed with bond lengths ranging from 2.0 to 2.05 $\mathrm{{\AA}}$.
In Rec with a 7-atom-cavity (Fig.~\ref{structure}b), the molecule locates at an fcc-hollow site of the second Ni layer.
Three C atoms of the C$_{6}$ hexagon bind to three Ni( "q" in Fig. \ref{structure}d) in the second layer and six C in the second layer of C$_{60}$ bind to six Ni("o") atoms in the top surface layer, forming  nine C$-$Ni bonds in all. The corresponding bond lengths rang from 1.93 to 2.00 $\mathrm{{\AA}}$, indicating stronger bonding strength than the one in Unrec.
The formation of more and stronger C$-$Ni bonds in Rec surpasses the energy cost of forming a seven-atom hole in the surface and makes the Rec structure more stable.
The C$_{60}$/Ni(111) system was investigated previously by STM \cite{c60_reconstruction_ni}, and two adsorption configurations were found. C$_{60}$ in one configuration was found to be 2.2 $\mathrm{{\AA}}$ lower than the other from the measured apparent height profile\cite{c60_reconstruction_ni}.
The C$_{60}$ height difference between Rec and Unrec in our calculation is 2.0 $\mathrm{{\AA}}$, by defining the molecular height as the height of the outermost hexagon, and using the average height of the second substrate layer's Ni atoms as a reference.
Thus, the low and high configurations in the STM experiment can be interpreted to be the Rec and Unrec structures in our calculation.
The coincidence in molecular height between theory and experiment supports the reliability of our proposed structures.

We quantify the spin-dependent charge transfer by the Bader charge analysis method\cite{bader}.
In Unrec, 0.64 electrons are donated into each C$_{60}$ from the metal substrate, among which 0.26 electrons occupy spin up states while the other 0.38 occupy spin down states.
The charge transfer increases to 1.82 electrons in Rec, with 0.85 electrons occupying spin up and 0.97 electrons occupying spin down.
These charge transfer have significant effects on the electronic and magnetic properties of both the adsorbed C$_{60}$ and the Ni surface, as detailed below.

\begin{figure}
\includegraphics[width=0.9\textwidth]{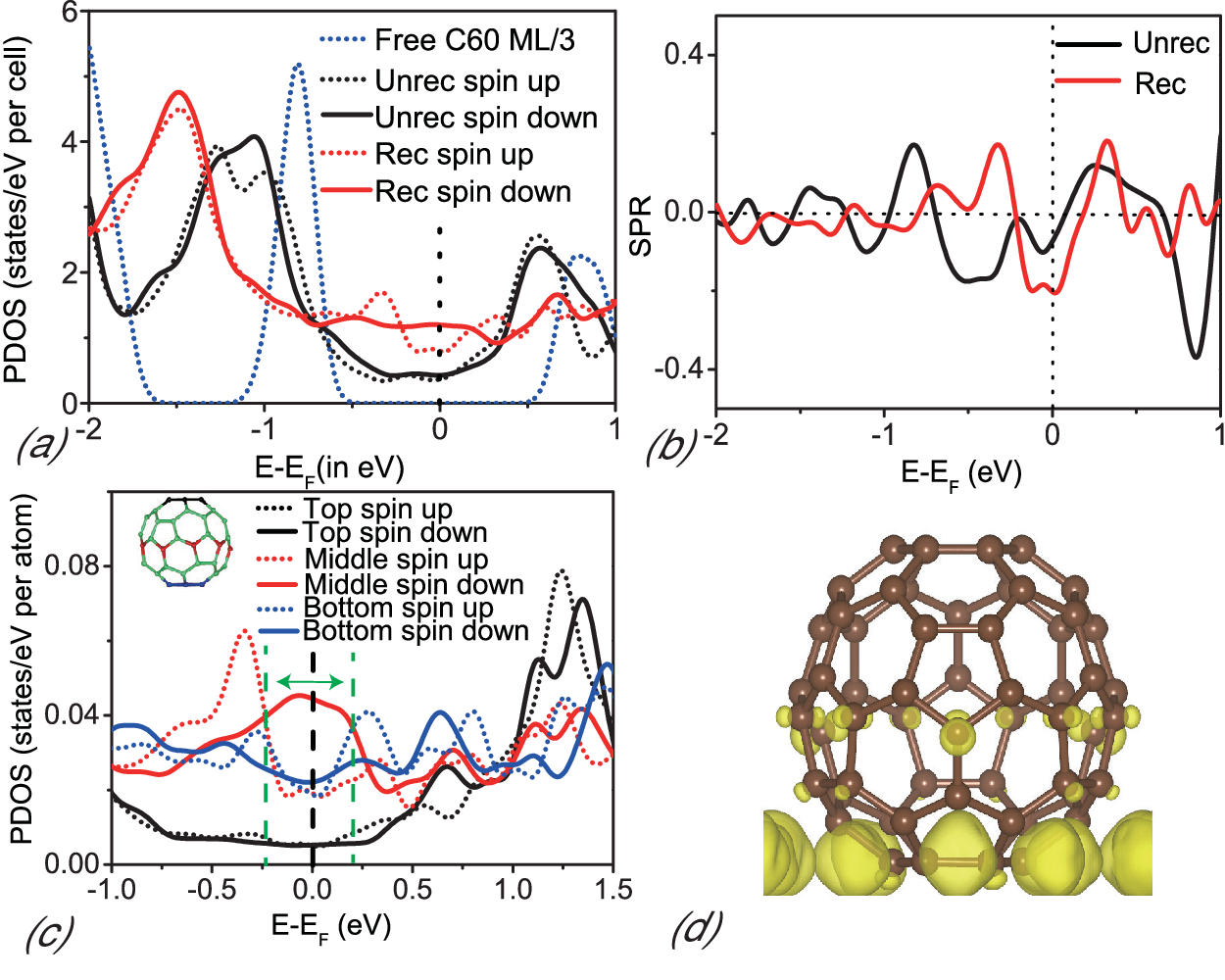}
\caption{\label{dos}(a) PDOS of free C$_{60}$ monolayer, and of C$_{60}$ in Rec and Unrec structures. (b) Spin-polarization ratio (SPR) of C$_{60}$ in Rec and Unrec.
(c) PDOSs of different parts of C$_{60}$ in Rec: the bottom (blue), the middle (red), and the top (black); the colors of PDOS curves match the colors of the atoms in the inserted C$_{60}$, the metal surface being horizontal and below in this view; the green dashed lines enclose the most suitable energy range for the molecule to filter spin.
(d) Spatial distribution of spin down local DOS in the energy interval of $[0, 0.2]$ eV (isovalue 0.0015 $e/Bohr^3$), viewed along the surface.}
\end{figure}

We plot the spin-polarized projected density of states (PDOS) of C$_{60}$ for Rec and Unrec in Fig.~\ref{dos}a, and the corresponding spin polarization ratio (SPR) in Fig.~\ref{dos}b. The SPR is defined as $\text{SPR}(E)=(\text{DOS}_\uparrow(E)-\text{DOS}_\downarrow(E))/(\text{DOS}_\uparrow(E)+\text{DOS}_\downarrow(E))$.
Fig.~\ref{dos}a shows that, compared to the free C$_{60}$ monolayer, the interaction with Ni broadens the molecular orbitals due to the hybridization of C$_{60}$ $p$ orbitals and substrate $d$ states\cite{spinterface4}.
These hybridization enhance the PDOS of C$_{60}$ near the Fermi level, changing the molecule from semiconducting to metallic.
In particular, the PDOS of C$_{60}$ at the Fermi level in Rec is 2.5 times that in Unrec.
Therefore, we can expect a higher conductance in Rec than in Unrec.
More importantly, due to the  magnetic surface, the molecular DOS is spin polarized after adsorption, especially  near Fermi energy.
According to the Julliere model of spin-dependent tunneling\cite{j_model}, the MR of the system is positively correlated with the SPR at the Fermi energy.
From Fig. \ref{dos}b, the SPR in Rec is about $19\%$ around the Fermi energy, almost three times that in Unrec, which is about $7\%$. Therefore, from the PDOS and SPR, we can expect that the surface reconstruction enhances the conductance and MR of C$_{60}$ simultaneously.
This feather meets the requirement of the miniaturization of read-head sensors, as we mentioned at the beginning\cite{point1}.

Another notable character of Rec is the DOS distribution in C$_{60}$. We plot the average PDOS in the bottom, the middle, and the top parts of C$_{60}$ (Fig. \ref{dos}c).
It can be seen that these PDOSs are quite different from each other.
These differences mean the delocalized molecular orbitals near the Fermi level in C$_{60}$ are broken into groups of localized orbitals. We plot the local DOS in the energy interval of $[0, 0.2]$ eV in Fig. \ref{dos}d to show the orbital distribution (the shape of the local DOS in $[-0.2, 0]$ eV is similar). We can see that the states are localized at the equator of the C$_{60}$.
In particular, from the middle PDOS in Fig. \ref{dos}c, PDOS at the C$_{60}$ equator has a rather strong intensity near the Fermi level, and that the corresponding spin polarization at the Fermi level is significantly stronger than in the other two parts of the molecule.
The PDOS at the equator has two advantageous features. One is that the Fermi level passes through the spin polarized peak, so these states can be accessed by a low bias voltage to reduce power dissipation. The other is that the polarized states are concentrated around the Fermi level so that a large current can be expected. Thus, the Rec system can hopefully be used as a basis of building high efficient molecular spin filters that work in the bias window of $\pm$0.2 eV, as is enclosed by green dashed lines in Fig. \ref{dos}c.
This feature is not present in the Unrec structure\cite{nic60ni}. The reason is as follows. In Unrec, only 0.6 electrons transfers into the three-fold degenerate lowest unoccupied molecular orbitals (LUMO) of C$_{60}$. Such a small electron transfer cannot significantly shift the C$_{60}$ LUMO toward the Fermi level (Fig.~\ref{dos}a). On the contrary, because of the reconstruction, 1.82 electrons transfer to C$_{60}$, and one LUMO orbital is supposed to be occupied. The partially occupied LUMO is the source of the states at the equator of C$_{60}$.
The layer-dependent DOS character of C$_{60}$ on Ni(111) is similar to the DOS character of a double-decker molecule after adsorption\cite{surface_exchange2}. The difference is that here the reconstruction plays a key role.

In addition to the above changes in the molecule, the reconstruction also significantly modifies the magnetic structures of the surface.
The calculated magnetic moments of the outermost nickel atoms in the Unrec or Rec cases are presented in Fig. \ref{structure}c and \ref{structure}d.
Based on the magnetic moment values and the internuclear distances from the molecule, the nickel atoms at the interface can be grouped into four types for both Rec and Unrec.
For the unreconstructed case, four Ni atoms are strongly affected ( "i" and "j" in Fig. \ref{structure}c): these four, which are right beneath the molecule in the outermost Ni layer, find their magnetic moments reduced to about 0.3 $\mu_{\mathrm{B}}$.
Next to these four atoms, four other Ni ("k") atoms are slightly influenced, converting their magnetic moments to 0.63 $\mu_{\mathrm{B}}$. Others (i.e. "l") keep their clean surface value of 0.69
$\mu_{\mathrm{B}}$.
By contrast, for the reconstructed case, all magnetic moments of the first Ni layer are significantly reduced. Six nickels ("o") along the rim of the hole change their magnetic moments to 0.28 $\mu_{\mathrm{B}}$ while three others ("p") change to about 0.44 $\mu_{\mathrm{B}}$.
As the reconstruction removes seven top layer atoms, the C$_{60}$ directly bonds to three nickels of the second layer ("q") and consequently reduces their magnetic moments to 0.35 $\mu_{\mathrm{B}}$.

\begin{table}
\caption{\label{magnetic_coupling}Magnetic coupling strengths $J_{\alpha\beta}{\mu}_{\alpha}{\mu}_{\beta}$ (in meV) between different types of mutually bonded atoms at the surface.
Here $\alpha, \beta =i, j, k, l$ for Unrec and  $\alpha, \beta =o, p, q, r$ for Rec indicate different types of atoms as labelled in Fig. \ref{structure}c and d.  }
\begin{ruledtabular}
\begin{tabular}{cccc}
Types(Unrec)&$J_{\alpha\beta}{\mu}_{\alpha}{\mu}_{\beta}$&Types(Rec)&$J_{\alpha\beta}{\mu}_{\alpha}{\mu}_{\beta}$\\ \hline
ii&1.14&oo&2.16\\
ij&1.21&op&1.92\\
ik&2.51&or&2.22\\
il&2.39&pr&1.08\\
jk&2.62&qq&1.95\\
jl&2.94&qr&4.68\\
kl&7.21&rr&6.97\\
ll&7.00& -& -  \\
\end{tabular}
\end{ruledtabular}
\end{table}

To further investigate the magnetic properties of the surface, we calculated the magnetic coupling and MAE of the surface atoms. We used a Heisenberg model with a Hamiltonian $H=-\sum_{\alpha \neq \beta}J_{\alpha\beta}\bm{\mu}_{\alpha}\cdot \bm{\mu}_{\beta}$ to describe the first nearest neighbor magnetic interaction between Ni atoms and defined the magnetic coupling strength between  atom $\alpha$ and atom $\beta$  as $J_{\alpha\beta}\mu_{\alpha}\mu_{\beta}$; $J_{\alpha\beta}\mu_{\alpha}\mu_{\beta}$ was calculated from energy differences between properly selected couples of magnetic configurations\cite{heisenberg_mapping}. For the clean unreconstructed surface, the coupling is 8.5 meV.
From Table~\ref{magnetic_coupling} we can see that the adsorption softens all the magnetic couplings between the surface atoms. This is similar with a previous study on Co interfaces\cite{surface_exchange3}. The decreases of these magnetic coupling strengths are strongly correlated with the distance from the molecule and the bonding condition to the molecule.

To analyze the reason of this softening, we examined the PDOS of the $d$ orbitals of selected Ni atoms\cite{SI}. These orbitals are grouped into $d_{\pi}$, which have out-of-surface-plane components ($d_{\pi}=d_{xz}+d_{yz}+d_{z^2}$), and $d_{\sigma}$ lie within the surface plane. We find a notable change is that the spin down orbitals move to lower energy and the spin up orbitals move oppositely so that the spin-split energy ($E_{SS}$) is reduced. We list the $d$-band-center shifts of the selected atoms relative to the clean unreconstructed surface and the corresponding decreases in $E_{SS}$ in Table \ref{dbands}.
We can see that the decrease of $E_{SS}^{d_{\sigma}}+E_{SS}^{d_{\pi}}$ on these atoms are qualitatively consistent with the reduction of the magnetic moments of the corresponding atoms, i.e. on the order of  $o\approx i>q>p$. Another feature is that the spin down orbitals shift relatively more in the Ni atoms which have C$-$Ni bonds. Thus, the C$-$Ni bonds play an important role in the orbital shifts. As is demonstrated in valence bond theory\cite{coordination_chemistry}, 3$d$ metals usually move some $d$ electrons of majority spin into its minority orbitals and use the empty $d$ orbitals to form new hybridized orbitals when forming bonds with organic compounds. So we conclude that the spin down orbital shifts are caused by hybridization. Meanwhile, as the substrate loses electrons, the center of the total $d$ orbitals must move to higher energy. Therefore, the spin up orbitals move toward the Fermi level. These are the reasons for the relative energy shifts of $d$ orbitals.

\begin{table}
\caption{\label{dbands} Energy shifts of $d$ band centers relative to the clean unreconstructed surface and of $E_{SS}$ for Ni$_{o}$, Ni$_{p}$, Ni$_{q}$ and Ni$_{i}$ (in eV) defined in Fig.~\ref{structure}. The reference $d$ band centers are $-$1.49, $-$0.81, $-$1.39, and $-$0.74 eV for d$_{\sigma}^{up}$, d$_{\sigma}^{down}$, d$_{\pi}^{up}$, and d$_{\pi}^{down}$, respectively.}
\begin{ruledtabular}
\begin{tabular}{ccccccc}
Atom types& $d_{\sigma}^{up}$ &$ d_{\sigma}^{down}$ & $E_{SS}^{d_{\sigma}}$ & $d_{\pi}^{up}$ & $d_{\pi}^{down}$ & $E_{SS}^{d_{\pi}}$ \\ \hline
Ni$_{o}$& 0.22 & $-$0.19 & 0.27	&0.14	&$-$0.21	&0.30 \\
Ni$_{p}$& 0.26 & $-$0.01 & 0.41	&0.21	&$-$0.03	&0.41 \\
Ni$_{q}$& 0.08 & $-$0.20 & 0.40	&$-$0.03&$-$0.36	&0.32 \\
Ni$_{i}$& 0.09 & $-$0.28 & 0.31	&$-$0.01&$-$0.41	&0.25 \\
Clean surface & 0  & 0  &  0.68 & 0 & 0 & 0.65 \\
\end{tabular}
\end{ruledtabular}
\end{table}

The MAE are calculated from the difference of spin-orbit interaction (SOI) energies for spins along different axes. The spin-orbit interaction can be calculated by non-collinear density functional theory (DFT) as well as a second order perturbation $E_{SOI}=-\lambda^2 \sum_{u,o}\frac{|\langle o|\mathbf{L}\cdot \mathbf{S}|u\rangle|^2}{E_u-E_o}$ \cite{perturbation1,perturbation2,perturbation3}, where $o$ and $u$ denote occupied and unoccupied collinear Kohn$-$Sham orbitals and $\lambda$ is a coupling constant.
The MAE of each layer can be obtained by linking the results of these two methods \cite{SI}. Values along the axes [111] (out of the surface plane), [101] and [112] (in plane, see Fig. \ref{structure}c) were obtained. The clean unreconstructed surface has an in-plane magnetization with an MAE of 0.24 meV/atom with degenerate energy for the easy axes [112] and [101]. After C$_{60}$ adsorption, the  MAE of the first layer becomes 0.14 meV/atom and 0.03 meV/atom in Unrec and Rec, respectively. This behavior coincides with the knowledge of the origin of MAE that the enhancement of vertical interaction (by molecule here) can weaken the stability of in-plane magnetization\cite{XCRD}.

From the above results, we can see that the adsorption of C$_{60}$ on Ni(111) will soften the magnetism of the outermost Ni atoms. Such softening is significantly enhanced by reconstruction. Therefore the reconstruction can be used as a convenient way to generate hard/soft composite magnetic structures to realize specific functions\cite{magnetic_spring}.

In summary, we determined the atomic structures at the C$_{60}$/Ni(111) with and without reconstruction.
The reconstruction not only stabilizes the molecule but also causes significant changes in the spintronic and magnetic properties at both sides of the organic/ferromagnetic interface.
On the molecule side, by reconstruction , the DOS and SPR of the adsorbed molecule at Fermi level are increased, which could simultaneously improve the conductance and MR of the system.
The reconstruction also creates a new spin-polarized and energy-concentrated state at the equator of C$_{60}$ near the Fermi energy, which makes it possible for the molecule to be used as a spin filter.
This suggests that, instead of using double-decker molecules, one can also use three-dimensional molecules, such as fullerenes, to obtain layer-dependent spin-polarized states.
On the surface side, reconstruction drastically reduces the exchange coupling and MAE of the outermost layer. The change of the exchange coupling can be related to the $d$-bands shift under the influence of molecular adsorption.
These findings reveal the importance of reconstruction on the organic/ferromagnetic interfaces, and could serve as basis for developing novel spintronic devices.
Besides, another our investigation found that C$_{60}$ could induce a different type of reconstruction on the ferromagnetic Fe(100)\cite{ c60_fe}. Combining this with the knowledge that C$_{60}$ induces non-magnetic metal surfaces reconstructions in different types, it is reasonable to believe that various reconstructions can also happen at other interfaces between different magnetic surfaces and molecules (e.g. C$_{70}$, thiolates, and graphene). Thus our discoveries can be extended to other systems with various combinations between organic materials and magnetic surfaces.

\begin{acknowledgments}
This work is supported in part by the National Natural Science Foundation of China (Grants 11474145 and 11334003). M. A. Van Hove was supported by the HKBU Strategic Development Fund.  We thank the National Supercomputing Center in Shenzhen for providing computation time, as well as the High Performance Cluster Computing Centre, Hong Kong Baptist University, which receives funding from the Research Grants Council, University Grants Committee of the HKSAR and Hong Kong Baptist University.
\end{acknowledgments}

\bibliography{mybib}

\end{document}